\documentclass[12pt]{amsart}
\usepackage{amsmath}
\usepackage{amssymb}
\usepackage[latin1]{inputenc}
\usepackage[T1]{fontenc}
\newtheorem{lemma}{Lemma}
\newtheorem{theorem}{Theorem}
\newtheorem{proposition}{Proposition}

\newcommand{\tr}{{\rm Tr }}

\newcommand{\aut}{{\rm Aut }}

\newcommand{\intd}[2]{D(#1,#2)}
\newcommand{\bra}{\langle}
\newcommand{\ket}{\rangle}
\newcommand{\vp}{\varphi}

\newcommand{\C}{\mathbb{C}}
\newcommand{\N}{\mathbb{N}}

\newcommand{\R}{\mathbb{R}}
\newcommand{\be}{\begin{equation}}
\newcommand{\eeq}{\end{equation}}
\newcommand{\bet}{\begin{equation*}}
\newcommand{\eeqt}{\end{equation*}}
\newcommand{\bea}{\begin{eqnarray}}
\newcommand{\eeqa}{\end{eqnarray}}
\newcommand{\beat}{\begin{eqnarray*}}
\newcommand{\eeqat}{\end{eqnarray*}}
\newcommand{\goesto}{\longrightarrow}
\newcommand{\h}[1]{\mathcal{#1}}
\newcommand{\hil}{\mathcal{H}}

\newcommand{\kil}{\mathcal{K}}

\newcommand{\hO}{\mathcal{O}}
\newcommand{\hT}{\mathcal{T}}

\newcommand{\hA}{\mathcal{A}}
\newcommand{\hB}{\mathcal{B}}

\newcommand{\hF}{\mathcal{F}}
\newcommand{\hK}{\mathcal{K}}
\newcommand{\intsqd}[2]{\widetilde{D}(#1,#2)}
\newcommand{\cc}[1]{\overline{#1}}

\begin{document}
\title{Notes on phase space quantization}
\author{J. Kiukas}
\address{Jukka Kiukas,
Department of Physics, University of Turku,
FIN-20014 Turku, Finland}
\email{jukka.kiukas@utu.fi}
\author{P. Lahti}
\address{Pekka Lahti,
Department of Physics, University of Turku,
FIN-20014 Turku, Finland}
\email{pekka.lahti@utu.fi}
\author{K. Ylinen}
\address{Kari Ylinen,
Department of Mathematics, University of Turku,
FIN-20014 Turku, Finland}
\email{kari.ylinen@utu.fi}
\begin{abstract}
We consider questions related to a quantization scheme in which a classical variable $f:\Omega\to \R$ on a phase space $\Omega$ is
associated with a semispectral measure $E^f$, such that the moment operators of $E^f$ are required to be
of the form $\Gamma(f^k)$, with $\Gamma$ a suitable mapping from the set of classical variables to the set of
(not necessarily bounded) operators in some Hilbert space. In particular, we investigate the situation where the map $\Gamma$ is implemented by the
operator integral with respect to some fixed positive operator measure. The phase space $\Omega$ is first taken to be an abstract
measurable space, then a locally compact unimodular group, and finally $\R^2$, where we determine explicitly the relevant operators
$\Gamma(f^k)$ for certain variables $f$, in the case where the quantization map $\Gamma$ is implemented by a translation covariant positive operator measure.
In addition, we consider the question under what conditions a positive operator measure is projection valued.
\end{abstract}
\maketitle

\section{Introduction}

Quantization can be any procedure which associates a quantum mechanical observable to a given
classical dynamical variable. The traditional way to realize a quantization is to assign to each classical variable
a Hermitean (symmetric, or even essentially selfadjoint) operator which should describe the quantum observable.
In the modern view of a quantum observable as a semispectral measure, this kind of quantization is no longer sufficient.
The question is how to modify the traditional scheme in order to fit it into the context of modern quantum mechanics. 

Classical variables can be represented by real valued measurable functions defined on some measurable space $(\Omega,\hA)$,
which is the phase space of the classical system.
The phase space can be taken to be e.g. $\R^{2n}$, in which case the variables
are Borel functions. In the conventional approach to quantization, we would have a map
$\Gamma$ from the set of real measurable functions to the set of
all linear (not necessarily bounded) operators in $\hil$, and $\Gamma(f)$ would be the observable corresponding to the classical variable $f$
(see e.g \cite{Dubin, Folland, Landsman, Schroeck, Stulpe}).
We would like to modify this scheme
so that we could assign a semispectral measure to the function $f$, instead of an operator. To do this, we use all the operators $\Gamma(f^k)$, $k\in\N$, instead
of just the first one of them, and then consider the moment problem of finding the unique semispectral measure
$E^f:\hB(\R)\to L(\hil)$ with the property that $\int x^k dE^f = \Gamma(f^k)$ for all $k\in \N$. Here $\int x^k dE^f$ is the operator integral
$L(x^k,E^f)$ of the function $x\mapsto x^k$ with respect to $E^f$; see the next Section for its definition. So if $\Gamma$ and $f$ are such that there
exists a unique solution $E^f$ of the described moment problem, then the collection of the operators
$\{ \Gamma(f^k)\mid k\in\N\}$ is eligible to represent a quantization $E^f$ of $f$.

We mention that the approach described above is used, for instance, in \cite{Wodkiewich} in the description of a quantum measurement.
In a typical measurement situation, where one aims to measure a (traditional) quantum observable represented by a selfadjoint operator,
one is actually measuring a noisy or unsharp version of that observable. The noisy version, represented by a semispectral measure,
may agree with the observable intended to be measured, on the statistical level of expectations. The unsharpness of the measurement
is reflected in the fact that the dispersion of the measurement statistics is actually greater than what would be obtained in the noiseless
case. The observable, as represented by a semispectral measure, cannot be described using a single operator. Accordingly, in the approach
of \cite{Wodkiewich}, each moment of the measurement outcome distribution is considered to be an average of certain
operator, called ''operational observable'', and the collection of these operators then represents the measured observable.
We want to point out that these ''operational observables'' are nothing else but the moment operators of the semispectral
measure representing the observable. Note that these moments need not determine the semispectral measure uniquely.
In certain special cases, however, they do \cite{LahtiIV, LahtiIII}, and sometimes even the first moment is enough \cite{DubinII, Pellonpää}.

One way to obtain a quantization map $\Gamma$ is to use the operator integral with respect to some given positive operator measure
$E$, i.e. define $\Gamma$ to be the map $f\mapsto L(f,E)$. In this case, if we define $E^f$ on the Borel sets of the real line by
$B\mapsto E(f^{-1}(B))$, we have simply $L(x^k, E^f)=L(f^k,E)$, as is easily seen by using the definition of the next Section and the usual change
of variables in the integral with respect to a complex measure. So then $E^f$ is a solution to the moment problem described above, which
leaves us with the uniqueness question. Note that if we choose $E$ to be a spectral measure, we end up getting only
spectral measures as the quantized observables, with all the operators $L(f^k,E)$ mutually commuting. We also remark that, in any case,
all the observables obtained via this type of quantization are \emph{functionally coexistent} \cite{Pulmannova}, so that they can be
measured together in the sense of Ludwig \cite[D.3.1, p. 153]{Ludwig}.

The structure of this note is as follows. In the Section 2 we give some results on the theory of operator integrals and
give a simple characterization of the quantization maps which can be represented by an
operator integral with respect to a positive operator measure. In Section 3, we consider the operator integral with respect to
covariant positive operator measures on a locally compact topological group.
Section 4 is devoted to the quantizations obtained by using the covariant phase space observables in $\R^2$, and
in Section 5, we discuss the ''optimal'' choice for such a covariant observable, in view of quantization. In the last Section,
we consider the question under what conditions a positive operator measure is a spectral one.

\section{The operator integral}

The basic tool in our quantization procedure is the operator integral. It associates a linear (not necessarily bounded)
operator in the Hilbert space $\hil$ of a quantum system to each complex measurable function on the phase space $\Omega$.
In this section we review the basic results concerning the theory of operator integrals and give some
additional remarks in relation to quantization.

Let $\Omega$ be a nonempty set and $\hA$ a $\sigma$-algebra of subsets of $\Omega$. Let $\hil$ be a complex Hilbert
space and $L(\hil)$ the set of bounded operators on $\hil$. Let $E:\hA\to L(\hil)$ be a positive operator measure,
i.e. a positive operator valued set function which is $\sigma$-additive with respect to weak
operator topology. For each $\vp,\psi\in\hil$, let $E_{\psi,\vp}$ denote the complex measure
$B\mapsto \bra\psi|E(B)\vp\ket$. We recall that a normalized positive operator measure $E$ is called a semispectral measure, and
that a semispectral measure $E$ is a spectral measure exactly when $E(A\cap B) = E(A)E(B)$ for all $A,B\in \hA$.
As mentioned in the Introduction, semispectral measures are also called observables.

Let $\hF(\Omega,\hA)$, or $\hF(\Omega)$ in brief, denote the set of all complex $\hA$-measurable functions defined on $\Omega$,
and $\hO(\hil)$ the set of all (not necessarily bounded) linear operators in $\hil$. For $f\in \hF(\Omega)$,
define
\bet
D(f,E)=\{ \vp\in\hil \mid f \text{ is } E_{\psi,\vp}\text{-integrable for each } \psi\in\hil \},
\eeqt
and
\bet
\intsqd{f}{E}=\{\vp\in\hil\mid |f|^2 \text{ is } E_{\vp,\vp}\text{-integrable}\}.
\eeqt
The following result was proved in \cite{Lahti}. The operator $L(f,E)$ appearing in it is called
\emph{the operator integral of $f$ with respect to E}.

\begin{theorem}\label{opintegral}
\begin{itemize}
\item[(a)] The set $D(f,E)$ is a linear (not necessarily dense) subspace of $\hil$, and there is a unique linear operator
$L(f,E)=\int f dE$ on the domain $D(f,E)$ satisfying
\bet
\bra\psi |L(f,E)\vp\ket = \int f dE_{\psi,\vp}
\eeqt
for all $\psi\in\hil$ and $\vp\in D(f,E)$.
\item[(b)] The set $\intsqd{f}{E}$ is a subspace of $D(f,E)$.
\item[(c)] If $f$ is real valued, $L(f,E)$ is a symmetric operator.
\item[(d)] While the inclusion $\intsqd{f}{E}\subset D(f,E)$ may in general be proper,
$\intsqd{f}{E}=D(f,E)$ in the case where $E$ is a spectral measure.
\end{itemize}
\end{theorem}
\noindent {\bf Remark. }
Since the operator measure $E$ is also strongly $\sigma$-additive, each set function
$\hA\ni B\mapsto E_{\vp}:=E(B)\vp\in\hil$, for $\vp\in\hil$, is an $\hil$-valued vector measure.
The definition of the operator integral states that $D(f,E)$ is the set of those $\vp\in\hil$ for which $f$ is integrable
with respect to the vector measure $E_{\vp}$ in the sense of \cite[p. 21]{Kluwanek}, and $L(f,E)\vp= \int f dE_{\vp}$
for each $\vp\in D(f,E)$. As pointed out in \cite[p. 37]{Kluwanek}, this definition of integrability is equivalent to
that of \cite[p. 323]{Dunford} (for proof, see \cite[Corollary 3.6]{Ylinen}).

\

The vector measure approach provides an easy way to characterize the operator integral $L(f,E)$
by approximating $f$ with bounded functions. For each $f\in \hF(\Omega)$, and $n\in\N$, let
$\tilde{f}_n$ be such that $\tilde{f}_n(x) = f(x)$ if $|f(x)|\leq n$, and $\tilde{f}_n(x)=0$ otherwise.
It is a well-known fact that in the case where $E$ is a spectral measure, we have
\bet
D(f,E) = \intsqd{f}{E}=D^{f,E} := \{\vp\in\hil \mid \lim_{n\goesto\infty} L(\tilde{f}_n,E)\vp\text{ exists}\}.
\eeqt
In addition, $L(f,E)\vp=\lim_{n\goesto\infty} L(\tilde{f}_n,E)\vp$ for all $\vp\in D^{f,E}$ (see e.g. \cite[p. 1196]{DunfordII}).
In the case of a general positive operator measure $E$, this need not be true.
For example, take a probability measure defined on the Borel sets of $\R$, such that it has a density which is an even function,
and $\int_0^\infty xd\mu(x) =\infty$. Let $E$ be the positive operator measure $B\mapsto \mu(B)I$, and $f(x)=x$.
Now $f$ is not $\mu$-integrable, so $D(f,E)=\{0\}$. But if $\vp\in\hil$, we have
$\bra \psi|L(\tilde{f}_n,E)\vp\ket = \bra \psi|\vp\ket\int_{-n}^n x d\mu(x)= 0$ for all $n\in \N$ and $\psi\in\hil$
because the density of $\mu$ is even, so that $D^{f,E} =\hil$.

Thus in the general case, the existence of the limit $\lim_{n\goesto\infty} L(\tilde{f}_n,E)\vp$ does not guarantee
that $\vp\in D(f,E)$. However, the following result holds.
\begin{proposition}\label{domainprop}
Let $E:\hA\to L(\hil)$ be a positive operator measure, and $f\in\hF(\Omega)$. Then
\bet
D(f,E) = \{\vp\in\hil \mid \lim_{n\goesto\infty} L(\chi_B\tilde{f}_n,E)\vp\text{ exists for each } B\in \hA\},
\eeqt
and $L(f,E)\vp=\lim_{n\goesto\infty} L(\tilde{f}_n,E)\vp$ for all $\vp\in D(f,E)$.
\end{proposition}
\begin{proof} Denote by $D$ the right hand side of the set equality appearing in the statement.
Let $\vp\in D$, and let $B\in \hA$, with $\eta_B = \lim_{n\goesto\infty} L(\chi_B\tilde{f}_n,E)\vp$.
Since each $\tilde{f}_n$ is bounded, we can choose
a sequence of $\hA$-simple functions $g_n$, such that $| g_n(x)-\tilde{f}_n(x)|\leq \frac 1n$ for each $n\in \N$
and $x\in \Omega$.
%(see [SKLIII2.2])
Clearly the sequence $(g_n)$ converges to $f$ pointwise. Now
\bet
\|L(\chi_Bg_n - \chi_B\tilde{f}_n, E)\vp\|= \sup_{\|\psi\|\leq 1} \left|\int_B (g_n-\tilde{f}_n)dE_{\psi,\vp}\right|
\leq \frac 1n \sup_{\|\psi\|\leq 1}|E_{\psi,\vp}|(\Omega)\leq \frac 4n \|E(\Omega)\|\|\vp\|,
\eeqt
so that
\bet
\|\int_B g_n dE_{\vp}-\eta_B\| \leq \frac 4n \|E(\Omega)\|\|\vp\|+\|L(\chi_B\tilde{f_n}, E)\vp-\eta_B\|.
\eeqt
It follows that the sequence $(\int_B g_n dE_{\vp})$ of vectors converges for each $B\in\hA$ (to $\eta_B$), so by the
definition of \cite[p. 323]{Dunford}, $f$ is integrable with respect to the vector measure $E_{\vp}$, i.e. $\vp\in D(f,E)$
(see the Remark following Theorem \ref{opintegral}).
Conversely, let $\vp\in D(f,E)$. Since $D(f,E) = D(|f|,E)$ by definition, and for each $B\in\hA$
the sequence $(\chi_B\tilde{f}_n)$ converges to $\chi_B f$ pointwise, with $|\chi_B\tilde{f}_n|\leq |\chi_Bf|$, it
follows e.g. from the dominated convergence theorem for the vector measure $E_{\vp}$ (see \cite[p. 328]{Dunford}) that
\bet
L(\chi_Bf, E)\vp= \int \chi_B f dE_{\vp} = \lim_{n\goesto\infty} \int \chi_B\tilde{f}_n dE_{\vp}
= \lim_{n\goesto\infty} L(\chi_B\tilde{f}_n,E)\vp
\eeqt
for each $B\in\hA$. Thus $\vp\in D$, and $L(f,E)\vp=\lim_{n\goesto\infty} L(\tilde{f}_n,E)\vp$.
\end{proof}

\noindent {\bf Remark. }
\begin{itemize} \item[(a)] We now have the subspace inclusions $\intsqd{f}{E} \subset D(f,E)\subset D^{f,E}$,
with each of them possibly proper. In the case where $E$ is a spectral measure, both inclusions are equalities.
\item[(b)] It is well known that in the case where $E$ is a spectral measure, the domain $D(f,E)$ is dense.
As seen before Proposition \ref{domainprop}, this need not be the case in general, so a question arises what is required for $E$ and
$f$ to make $D(f,E)$ dense.
\item[(c)] Another difference to the spectral case is that for real valued $f\in\hF(\Omega)$, the symmetric operator
$L(f,E)$ is not necessarily selfadjoint (the adjoint need not even exist), and it seems to be
difficult, in general, to determine when $L(f,E)$ might have selfadjoint extensions. In the case where $f$ is positive
and $D(f,E)$ dense, the positive symmetric operator $L(f,E)$ of course has its selfadjoint Friedrichs extension.
\end{itemize}

\

As noted in the Introduction, the starting point of our quantization scheme is a map $\Gamma$ from real valued
measurable functions to the set $\hO(\hil)$.
The following theorem characterizes those maps $\Gamma:\hF(\Omega)\to \hO(\hil)$ which are
implemented by an operator integral. The corresponding result involving bounded functions is well known.
Here the functions $\tilde{f}_n$ are defined for each $f$ as in Proposition \ref{domainprop}.
\begin{theorem}\label{theorem2}
A map $\Gamma:\hF(\Omega)\to \hO(\hil)$ coincides with the map $f\mapsto L(f,E)$ for a (clearly unique) positive operator
measure $E$ if and only if the following conditions are satisfied.
\begin{itemize}
%\item[(i)] $a \Gamma(f)+b \Gamma(g) \subset \Gamma(a f+b g)$ for $a,b\in\R$ and $f,g\in \hF(\Omega)$;
\item[(i)] $\Gamma$ restricted to bounded functions is a positive linear map with values in $L(\hil)$;
\item[(ii)] if $(f_n)$ is an increasing sequence of positive $\hA$-measurable functions converging pointwise to a bounded
$f\in\hF(\Omega)$, then $\sup_{n\in\N} \bra \vp |\Gamma(f_n) \vp\ket = \bra \vp|\Gamma(f)\vp\ket$ for each $\vp\in\hil$;
\item[(iii)] for each $f\in\hF(\Omega)$, the domain $D(\Gamma(f))$ of $\Gamma(f)$ consists of those vectors
$\vp\in \hil$ for which the sequence $(\Gamma(\chi_B\tilde{f}_n)\vp)$ of vectors converges for each $B\in \hA$.
\item[(iv)] for each $\vp\in D(\Gamma(f))$, the sequence $(\Gamma(\tilde{f}_n)\vp)$ converges to $\Gamma(f)\vp$.
\end{itemize}
\end{theorem}
\begin{proof} Assume first that there is a positive operator measure $E$, such that $\Gamma(f) = L(f,E)$
for each $f$. The above properties follow easily:
%The property (i) is a consequence of the linearity of the integral with respect to a complex measure
%$E_{\psi,\vp}$, and the triangle inequality $|a f+b g|\leq |a| |f|+|b||g|$.
The property (i) is well known (see e.g. \cite[pp. 26-28]{Berberian}), and (ii)
follows from the monotone convergence theorem. Proposition \ref{domainprop} gives (iii) and (iv).
Next assume that (i)-(iii) hold for a map $\Gamma:\hF(\Omega)\to \hO(\hil)$.
By (i) the map $\hA\ni B\mapsto E^{\Gamma}(B) := \Gamma(\chi_B)\in \hO (\hil)$ is a positive operator valued
additive set function. Since $\Gamma$ is positive for bounded functions by (i), $\sup_{n\in\N}$ in condition of (ii)
can be replaced by $\lim_{n\goesto \infty}$. This implies that each set function
$B\mapsto \bra \vp|E^{\Gamma}(B)\vp\ket$
is a positive measure, so that $E^{\Gamma}$ is a positive operator measure. If $f\in \hF(\Omega)$ is a bounded
positive function, we have $\bra \vp|\Gamma(f)\vp\ket = \bra \vp|L(f,E^{\Gamma})\vp\ket$ for all
$\vp\in \hil$, as is seen by approximating $f$ with an increasing sequence of simple functions and using linearity,
(ii), and the monotone convergence theorem. Hence, if $f\in \hF(\Omega)$ is bounded, it follows by
linearity and polarization that $\Gamma(f) = L(f,E^{\Gamma})$. Now let $f\in\hF(\Omega)$ be arbitrary. It follows by (iii)
and Proposition \ref{domainprop} that $D(\Gamma(f)) = D(f,E^{\Gamma})$, and (since each $\tilde{f}_n$ is bounded), also
\bet
L(f,E^{\Gamma})\vp = \lim_{n\goesto\infty} L(\tilde{f}_n, E^{\Gamma})\vp =
\lim_{n\goesto\infty} \Gamma(\tilde{f}_n)\vp = \Gamma(f)\vp
\eeqt
for all $\vp\in D(\Gamma(f))$ (where (iv) is used), so $L(f,E^{\Gamma}) = \Gamma(f)$.
\end{proof}
\noindent {\bf Remark. } In the quantization scheme described in the Introduction the classical variables were thought to be real valued.
Obviously, the preceding Theorem holds also with $\hF(\Omega)$ replaced by the set of real $\hA$-measurable functions.

\

We end this section by discussing briefly a simple way of obtaining quantization maps without the use of positive operator
measures. This approach is essentially the one frequently used in the conventional quantization (e.g. Weyl quantization):
the operator corresponding to a given classical variable $f:\R^{2n}\to \R$ is obtained by integrating the variable with respect to some operator
valued function defined on the phase space. The quantization of the variable then becomes a continuous distribution valued
operator defined on some dense subspace of $L^2(\R)$, which does not depend on the variable itself. In the case of Weyl quantization, for
example, all the quantized operators are defined on
a common domain (see e.g. \cite{Dubin, Voros}). However, to make the situation similar to that of the operator integral map considered above,
we define the quantization map in the following simple way. The proof is a direct adaptation of the proof of Theorem \ref{opintegral} (a)
(see \cite{Lahti}). Note that it follows from (i) that the integrand in (ii) is $\hA$-measurable. This is so because
$\|\vp\|=\sup \{|\bra \psi|\vp\ket|\mid \psi\in \h M, \|\psi\|=1\}$ for any vector $\vp\in \hil$, where $\h M$ is a fixed countable dense set in the
separable Hilbert space $\hil$. Hence the integral in (ii) is well defined.

\begin{proposition} Let $(\Omega,\hA,\mu)$ be a measure space, $\hil$ a separable Hilbert space, and $\Lambda:\Omega\to L(\hil)$
a map with the following properties:
\begin{itemize}
\item[(i)] $\omega \mapsto \bra \psi|\Lambda(\omega)\vp\ket$ is $\hA$-measurable for all $\psi,\vp\in\hil$;
\item[(ii)] $\int_B \|\Lambda(\omega)\vp\|d\mu(\omega)<\infty$ for each $\vp\in \hil$ and $B\in \hA$ with $\mu(B)<\infty$.
\end{itemize}
Then for each $\hA$-measurable function $f:\Omega\to \C$, there exists a linear operator $\Gamma_{\Lambda}(f)$ in $\hil$,
such that
\bet
D(\Gamma_{\Lambda}(f)) =\{ \vp\in\hil \mid f\bra \psi|\Lambda(\cdot)\vp\ket \text{ is } \mu\text{-integrable for each } \psi\in \hil \},
\eeqt
and
\bet
\bra \psi |\Gamma_{\Lambda}(f)\vp\ket = \int f(x) \bra \psi|\Lambda(\omega)\vp\ket d\mu(\omega) \ \  \psi\in \hil,\vp\in D(\Gamma_{\Lambda}(f)).
\eeqt
\end{proposition}
\begin{proof}
It is clear that $D(\Gamma_{\Lambda}(f))\subset\hil$ is a vector subspace. Let $(f_n)$ be a sequence of simple functions converging pointwise to $f$,
with $|f_n|\leq |f|$ for all $n\in\N$. Since $|\bra \psi |\Lambda(\omega)\vp\ket|\leq \|\psi\| \|\Lambda(\omega)\vp\|$ for each $\psi,\vp\in\hil, \omega\in \Omega$,
it follows from (ii) and (i) that for each $n\in \N$, we have $D(\Gamma_{\Lambda}(f_n))=\hil$ and the linear functional
$\psi \mapsto \int \cc{f_n(\omega) \bra \psi|\Lambda(\omega)\vp\ket} d\mu(\omega)$ is continuous for each $\vp\in\hil$. Hence, for each $n\in \N$ and $\vp\in \hil$, there is
$\eta_n^{\vp}\in \hil$, such that
\bet
\bra \psi|\eta_n^{\vp}\ket = \int f_n(\omega) \bra \psi|\Lambda(\omega)\vp\ket d\mu(\omega)
\eeqt
for all $\psi\in\hil$. Now, let $\vp\in D(\Gamma_{\Lambda}(f))$. Since $|f_n|\leq |f|$ for all $n$, the dominated convergence theorem implies that
the sequence $\bra \psi|\eta_n^{\vp}\ket$ converges for each $\psi\in \hil$ to $\int f(\omega) \bra \psi|\Lambda(\omega)\vp\ket d\mu(\omega)$, so by the
uniform boundedness theorem and the reflexivity of $\hil$, there is $\Gamma_{\Lambda}(f)\vp\in\hil$, such that
\bet
\bra \psi |\Gamma_{\Lambda}(f)\vp\ket = \int f(\omega) \bra \psi|\Lambda(\omega)\vp\ket d\mu(\omega)
\eeqt
for all $\psi \in \hil$. Clearly, the map $D(\Gamma_{\Lambda}(f))\ni \vp\mapsto \Gamma_{\Lambda}(f)\vp\in \hil$ is linear, so the proof is complete.
\end{proof}
\noindent {\bf Remark. } Consider the situation where we have a locally compact unimodular topological group $G$, with a
left Haar measure $\mu$, and a strongly continuous projective unitary representation $U:G\to \h U(\hil)$. Then, for any $A\in L(\hil)$,
the map $\Lambda:G\to L(\hil)$, defined by $\Lambda(g) = U(g)AU(g)^*$, satisfies the conditions of the preceding Proposition, so
we get the corresponding quantization map $\Gamma_{\Lambda}$. Notice that in the case where the representation
is square integrable and $A$ positive, the quantization map $\Gamma_{\Lambda}$ can be represented by an operator integral
if and only if $A$ has finite trace. Namely, the case $\tr[A]=\infty$ gives $D(\Gamma_{\Lambda}(\chi_{G})) = \{0\}$ (see e.g. \cite[Lemma 2]{wernerarticle}), while
the case $\tr[A]<\infty$ leads to the usual quantization map given by the operator integral with respect to a covariant positive operator
measure (multiplied by some constant).

Consider the case where $G=\R^2$, $U(g)=U((q,p))$ are the Weyl operators, and $A$ is the parity operator $L^2(\R)\ni \psi\mapsto \psi(-\cdot)\in L^2(\R)$
(multiplied by a suitable constant).
Now $\Gamma_{\Lambda}$ is the Weyl quantization map. It is well known that for $f\in L^1(\R^2)\cup L^2(\R^2)$, the operator
$\Gamma_{\Lambda}(f)$ is bounded. Moreover, if $f\in L^2(\R^2)$, then $\Gamma_{\Lambda}(f)$ is a Hilbert-Schmidt operator, and if
$f$ is a Schwartz function, then $\Gamma_{\Lambda}(f)$ is a trace class operator (see e.g. \cite{Voros}).
It is a well known fact that the Weyl quantizations of the classical position and momentum variables are the position and momentum
operators $Q$ and $P$, in the ''distributional'' sense. Note, however, that the actual domains of $Q$ and $P$ are not given by the formula
of the preceding theorem. For example, the characteristic function $\chi_{[-1,1]}$ is in the domain of $Q$, but
the function $(q,p)\mapsto q\bra \chi_{[-1,1]}|\Lambda(q,p)\chi_{[-1,1]}\ket$ is not (Lebesgue)-integrable, as is easily seen by calculating
the explicit form of the function. 

\section{Covariant quantization}

We now take the set $\Omega$ of the preceding section to be a locally
compact second countable unimodular topological group, henceforth denoted as $G$, and let $\hB(G)$ denote the
Borel $\sigma$-algebra of $G$. Fix $\lambda$ to be a Haar measure in $G$. Let $\hT(\hil)$ denote the Banach space of trace class
operators on the Hilbert space $\hil$, and let $\aut(\hT(\hil))$ denote the set
of linear, positive, trace norm preserving bijections from $\hT(\hil)$ onto itself. We consider it equipped with the topology given by
the functionals $\aut(\hT(\hil))\ni \beta\mapsto \tr[A\beta(T)]\in \C$, where $A\in L(\hil)$ and $T\in \hT(\hil)$.

Assume further that there is a continuous group homomorphism $\beta:G\to \aut(\hT(\hil))$ and
a constant $d>0$, satisfying
\be\label{squareintrep}
\int \tr[P_1\beta(g)(P_2)] d\lambda = d \  \ \text{ for all one-dimensional projections } P_1,P_2 \text{ on } \hil.
\eeq

We now consider quantizations connected to the structure of $G$ given by the homomorphism $\beta$, in
the following sense: A map $\Gamma:\hF(G)\to \hO(\hil)$ with the property that $\Gamma(f)\in L(\hil)$ for all bounded functions $f\in\hF(G)$
is said to be \emph{$\beta$-covariant}, if $\beta(g)^*(\Gamma(f))= \Gamma(f(g\cdot))$ for all $g\in G$ and
all bounded functions $f\in \hF(G)$.

If $\Gamma$ is such that it can be represented by the operator integral with respect to an observable
$E$, (i.e. $\Gamma$ satisfies the conditions of Theorem \ref{theorem2}), then it is straightforward to
verify that $\Gamma$ is $\beta$-covariant if and only if the observable $E$ is $\beta$-covariant in the following sense:
An observable $E:\hB(G)\to L(\hil)$ is said to be \emph{$\beta$-covariant} if
$\beta(g)^*(E(B)) = E(g^{-1} B)$ for all $g\in G$, $B\in \hB(G)$.

Covariant observables are essential in quantum
mechanics, and hence they have been studied quite extensively. The canonical examples of covariant observables are constructed e.g. in \cite{Davies}, and
there are (at least) two completely different ways to obtain their characterization: a direct approach \cite{Holevo,WernerII,wernerarticle}, which uses
the theory of integration with respect to vector measures, and a group theoretical approach \cite{Cassinelli}. The most general of these
characterizations is in \cite{Cassinelli}. In our context of a unimodular group, the characterization is given by the following theorem.

\begin{theorem} Let $T$ be a positive operator of trace one. Then there is a $\beta$-covariant observable $E^T:\hB(G)\to L(\hil)$,
such that
\be\label{generated}
E^T(B) = d^{-1}\int_B \beta(g)(T) d\lambda(g)
\eeq
in the ultraweak sense for each $B\in \hB(G)$.
Conversely, assume that $E:\hB(G)\to L(\hil)$ is a $\beta$-covariant observable. Then there is a unique positive operator $T$
of trace one, such that $E=E^T$.
\end{theorem}
If $E$ is a $\beta$-covariant observable, we call the corresponding trace-one positive operator $T$ the \emph{generating operator} for $E$.
Thus, for a $\beta$-covariant observable $E$, we have
\bet
D(f, E) = \{ \vp\in \hil \mid g\mapsto f(g)\bra \psi|\beta(g)(T)\vp\ket \text{ is } \lambda\text{-integrable for each } \psi\in\hil \}
\eeqt
and
\bet
\bra \psi |L(f,E)\vp\ket = d^{-1}\int f(g) \bra \psi |\beta(g)(T)\vp\ket d\lambda(g)
\eeqt
for all $\vp\in D(f,E)$ and $\psi\in\hil$, where $T$ is the generating operator for $E$.

According to the Wigner theorem, each $\beta(g)$ has the form $\beta(g)(T) = U(g)TU(g)^*$ for some unitary or
antiunitary operator $U(g)$, which is unique up to a phase factor, so that $\beta(g)$ corresponds to the associated equivalence class
of unitary operators (see e.g. \cite[p. 19]{Symmetry} or \cite[p. 22]{Holevo}). It follows that, in the case where $G$ is connected, the map
$g\mapsto U(g)$ is a weakly (Borel) measurable projective unitary representation of $G$, where each $U(g)$ is chosen from the equivalence class
corresponding to $\beta(g)$ by means of some (measurable) section (see \cite[p. 23]{Holevo} and \cite[pp. 30, 100]{Symmetry}).
The relation \eqref{squareintrep} gives the so-called square integrability condition
\bet
\int |\bra \psi|U(g)\vp\ket|^2 d\lambda(g) = d
\eeqt
for all unit vectors $\psi,\vp\in\hil$. Clearly, for each $\vp,\psi\in \hil$ there is a $g\in G$ such that
$\bra \psi|U(g)\vp\ket \neq 0$. This implies that the closed linear span of $\{U(g)\vp\mid g\in G\}$ is dense in
$\hil$ for each $\vp\in\hil$, which means that the projective representation $g\mapsto U(g)$ is irreducible.
The $\beta$-covariance condition for an observable $E$ takes the form
$U(g)^*E(B)U(g) = E(g^{-1} B)$ for all $g\in G$, $B\in \hB(G)$.

\

Hence, we know that each covariant quantization map $\Gamma:\hF(G)\to \hO(\hil)$, which can be represented by an operator integral, is of the form
$\Gamma = \Gamma^T:=L(\cdot,E^T)$ for some generating operator $T$.
It is worth noting that in certain cases the observables produced by the quantization scheme associated with a map
$\Gamma^T$ are never spectral measures. Namely, the irreducibility of $U$ implies the following, perhaps well-known result.

\begin{proposition}\label{projections} Assume that $G$ is connected, and that the projective representation $U$ associated with $\beta$ is strongly continuous.
Let $T\in \hT(\hil)$ be positive and of trace one. Then the only projections in the range of $E^T$ are $O$ and $I$.
\end{proposition}
\begin{proof}
First we notice that if $E^T(X)$ is a projection for some $X\in \hB(G)$ and positive operator $T$ of trace one, then there is a nonzero $\vp\in\hil$,
such that $E^{|\vp\ket\bra\vp|}(X)$ is a projection. Indeed, let $T$ be a positive operator of trace one, $\lambda>0$ an eigenvalue
of $T$ (so that $\lambda \leq 1$) and $\vp\in\hil$ an associated eigenvector. Then we can decompose $T$ as
$T= \lambda |\vp\ket\bra\vp| +(1-\lambda)T'$, where $|\vp\ket\bra\vp|$ and $T'$ are positive and of trace one, so we can write
$E^T(X) = \lambda E^{|\vp\ket\bra\vp|}(X)+(1-\lambda) E^{T'}(X)$. Since any projection is an extreme point of the convex set
$\{A\in L(\hil) \mid 0\leq A\leq I\}$ (see e.g. \cite[p. 19]{Davies}), it follows that if $E^T(X)$ is a projection, then $E^T(X)=E^{|\vp\ket\bra\vp|}(X)$.

Hence, it suffices to show that for each unit vector $\eta\in\hil$, the only projections in the range of $E^{|\eta\ket\bra\eta|}$ are $O$ and $I$.
Denote $T=|\eta\ket\bra\eta|$, and assume that there is a projection $P$ in the range of $E^T$. Then $PE^T(B)=E^T(B)P$ for all
$B\in \hB(G)$ \cite{LahtiV}. Let $\vp\in \hil$. Now
\bet
\int_B \bra \vp |P\beta(g)(T)\vp\ket d\lambda(g) = \int_B \bra \vp |\beta(g)(T)P\vp\ket d\lambda(g)
\eeqt
for all $B\in \hB(G)$. Since $g\mapsto \bra \vp |(P\beta(g)(T) - \beta(g)(T)P)\vp\ket$ is continuous, it is thus zero
for all $g\in G$. Hence, $P\beta(g)(T) = \beta(g)(T)P$ for all $g\in G$.

Let $U_{\eta}$ denote the map $g\mapsto U(g)\eta$. We then have
\bet
P|U_{\eta}(g)\ket\bra U_{\eta}(g)| = |U_{\eta}(g)\ket\bra U_{\eta}(g)|P
\eeqt
for all $g\in G$.

It follows that for each $g\in G$, either $U_{\eta}(g)\in P(\hil)$
or $U_{\eta}(g)\in P(\hil)^{\perp}$. Let $f:G\to \{0,1\}$ be the function such that $f(g) = 0$ if $U_{\eta}(g)\in P(\hil)$ and $f(g) = 1$ if $U_{\eta}(g)\in P(\hil)^{\perp}$.
Then $f$ is continuous, when the set $\{ 0,1\}$ is equipped with the discrete topology. Indeed, let $g_0\in G$. Since $U_{\eta}$ is continuous, 
$W=U_{\eta}^{-1}(\{ \vp\in \hil \mid \|U_{\eta}(g_0)-\vp\|< \sqrt{2}\})$ is an open set in $G$ containing $g_0$. Assume first that $f(g_0)=0$.
Since all vectors $U_{\eta}(g)$ are of unit length,
it follows that $\|U_{\eta}(g) -U_{\eta}(g_0)\|=\sqrt{2}$ whenever $f(g) = 1$. Hence, $f(W)\subset \{ 0 \}$. Similarly, if we assume that $f(g_0)=1$, it follows that
$f(W)\subset \{ 1\}$. This implies that $f$ is continuous. Since $G$ is connected, $f$ cannot be a surjection, so either $U_{\eta}(g)\in P(\hil)$ for all $g\in G$, or
$U_{\eta}(g)\in P(\hil)^{\perp}$ for all $g\in G$. But, due to the irreducibility of the projective representation $U$, the closed linear span of the set
$\{ U_{\eta}(g)\mid g\in G\}$ is dense in $\hil$. This is clearly possible only if either $P=I$ or $P=O$.
The proof is complete.
\end{proof}

The following observation is another consequence of the irreducibility of the projective representation associated with $\beta$.
It uses a calculation similar to that appearing e.g. in \cite[p.40]{Wong} in a different context. Part (a) is mentioned
also in \cite{Werner}.
 
\begin{proposition}\label{densityprop} Assume that $G$ is connected and let $U$ be the projective
representation associated with $\beta$. Let $E:\hB(G)\to L(\hil)$ be a $\beta$-covariant observable.
\begin{itemize}
\item[(a)] Assume that $f\in \hF(G)$ is such that $\intsqd{f}{E}\subset \intsqd{f(g\cdot)}{E}$ for all $g\in G$. Then
$U(g)\intsqd{f}{E} = \intsqd{f}{E}$ for all $g\in G$, and either $\intsqd{f}{E}=\{0\}$ or $\intsqd{f}{E}$ is dense.
\item[(b)] Assume that $f\in \hF(G)$ is such that $D(f,E)\subset D(f(g\cdot),E)$ for all $g\in G$. Then
$U(g)D(f,E) = D(f,E)$ for all $g\in G$, and either $D(f,E)=\{0\}$ or $D(f,E)$ is dense. Moreover,
\be\label{intcov}
U(g)^*L(f,E)U(g)\subset L(f(g\cdot),E)
\eeq
for all $g\in G$.
\end{itemize}
\end{proposition}
\begin{proof}
Let $T$ be the positive trace one operator associated with $E$, so that
\beat
\intsqd{f}{E} &=& \{ \vp\in \hil \mid g\mapsto |f(g)|^2 \bra\vp| U(g)TU(g)^*\vp\ket \text{ is } \lambda \text{-integrable }\};\\
D(f,E) &=& \{ \vp\in \hil \mid g\mapsto |f(g)| |\bra\psi| U(g)TU(g)^*\vp\ket| \text{ is }
\lambda \text{-integrable for all } \psi\in\hil\}.
\eeqat
For all $h, g\in G$, we have $U(h)^*U(g) = c(g,h) U(h^{-1}g)$, where $(h,g)\mapsto c(h, g)$
is some torus valued function, so that $U(g)^*U(h) = [U(h)^*U(g)]^* =c(g,h)^{-1} U(h^{-1}g)^*$, and hence
\be\label{UhUg}
U(h)^*U(g)TU(g)^*U(h)= U(h^{-1}g)TU(h^{-1}g)^*.
\eeq

(a) Let $\vp\in \intsqd{f}{E}$, and $h\in G$. By the left invariance of the Haar measure and (\ref{UhUg}), we have
\beat
\int |f(g)|^2 \bra U(h)\vp| U(g)TU(g)^*U(h)\vp \ket d\lambda(g)
&=& \int |f(g)|^2 \bra \vp| U(h^{-1}g)TU(h^{-1}g)^*\vp \ket d\lambda(g)\\
&=& \int |f(hg)|^2 \bra \vp| U(g)TU(g)^*\vp \ket d\lambda(g),
\eeqat
with all the integrands positive. Since $\vp\in \intsqd{f(h\cdot)}{E}$ by assumption, the last integral is finite, so
$U(h)\vp\in \intsqd{f}{E}$. Thus $\intsqd{f}{E}$ is an invariant subspace of the projective representation $U$,
implying that the closure $\cc{\intsqd{f}{E}}$ is a closed invariant subspace of $U$. It follows from the
irreducibility of $U$ that $\intsqd{f}{E}$ is either trivial or dense.
The fact that $U(h)\intsqd{f}{E} = \intsqd{f}{E}$ follows because we have $U(h^{-1}) = c'(h)U(h)^*$ for some
torus valued function $c'$. The proof of (a) is complete.

(b) Let $h\in G$, $\vp\in D(f,E)$, and $\psi\in\hil$. Then by using (\ref{UhUg}) and the assumption, we get
\beat
\int |f(g)| |\bra \psi| U(g)TU(g)^*U(h)\vp \ket| d\lambda(g)
&=& \int |f(g)| |\bra U(h)U(h)^*\psi| U(g)TU(g)^*U(h)\vp \ket| d\lambda(g)\\
&=& \int |f(g)| |\bra U(h)^*\psi| U(h^{-1}g)TU(h^{-1}g)^*\vp \ket| d\lambda(g)\\
&=& \int |f(hg)| |\bra U(h)^*\psi| U(g)TU(g)^*\vp \ket| d\lambda(g) <\infty,
\eeqat
so that $U(h)\vp\in D(f,E)$. Thus $D(f,E)$ is an invariant subspace for $U$, and hence $D(f,E)$ is either trivial or
dense. The fact that $U(h)D(f,E) = D(f,E)$ follows for the same reason as the corresponding one in (a).

Let $h\in G$. By repeating the preceding calculation without the absolute value signs, we get 
\bet
\bra \psi |L(f,E)U(h)\vp\ket = \bra U(h)^*\psi |L(f(h\cdot), E)\vp\ket = \bra \psi |U(h)L(f(h\cdot),E)\vp\ket
\eeqt
for each $\psi\in\hil$ and $\vp\in U(h)^*D(f,E) = D(f,E)\subset D(f(h\cdot),E)$, so that (\ref{intcov}) holds.
\end{proof}
\noindent {\bf Remark.}
\begin{itemize}
\item[(a)] If we assume $D(f,E)= D(f(g\cdot),E)$ for all $g\in G$, then Proposition \ref{densityprop}
(b) gives the strict operator equality $U(g)^*L(f,E)U(g) = L(f(g\cdot),E)$ for each $g\in G$, with dense domain $D(f,E)$.
This resembles the covariance condition for the observable $E$.
\item[(b)] Since each $E_{\psi,\vp}$ is a finite measure, it is clear that the conditions of Proposition
\ref{densityprop} (a) and (b) are satisfied e.g. by all functions
$f\in\hF(G)$ with the property that for each $h\in G$ there are nonnegative constants $K_h$ and $M_h$,
such that $|f(hg)|\leq K_h|f(g)|+M_h$ for almost all $g\in G$. In the case where $G=\R^{2n}$ (see the
beginning of the next Section), all polynomials of the form $\R^{2n}\ni (x_1,\ldots,x_{2n})\mapsto p(x_i)\in \R$, where
$p:\R\to\R$ is a polynomial and $i=1,\ldots,2n$, are like this.
\end{itemize}

\section{Phase space quantization on $\R^2$}

Consider the special case where $G=\R^{2}$, with $\lambda$ the Lebesgue measure. Fix $\{|n\ket\}$ to be an orthonormal basis of $\hil$, and
let $U:L^2(\R)\to \hil$ be the unitary operator which maps
the $n$th Hermite function $h_n$ to $|n\ket$.
Define $W(q,p) = UW_0(q,p)U^{-1}$, where $(W_0(q,p)f)(t) = e^{i\frac 12 qp} e^{ipt}f(t+q)$, and
$\beta:\R^2\to \aut (\hT(\hil))$ by $\beta(q,p)(T) = W(-q,p)TW(-q,p)^*$.
Now $\beta$ is a continuous qroup homomorphism, satisfying \eqref{squareintrep}, with $d=2\pi$, and
$\lambda$ the Lebesgue measure of $\R^2$.
Let $A_{\pm}$ be the ladder operators associated with the basis $\{|n\ket\}$, and define $Q$ and $P$ to be the closures of
the operators $\frac{1}{\sqrt{2}} (A_++A_-)$ and $\frac{1}{\sqrt{2}} i(A_+-A_-)$, respectively. Then $A_+ =A_-^*$, and
$Q$ and $P$ are unitarily equivalent to the position and momentum operators in $L^2(\R)$ via $U$.
Let $N$ denote the selfadjoint operator $A_+A_-$.

According to the general result described above, each positive
operator $T$ of trace one generates the map $f\mapsto L(f,E^T)$, where
\bet
E^T(B) = d^{-1}\int_B \beta(q,p)(T) d\lambda(q,p).
\eeqt
The generating operators $T$ of the form $T=\sum_n w_n |n\ket\bra n|$, where $\sum_n w_n =1$, and $w_n\geq 0$ for each $n$, have a special significance, as
they are the ones for which $E^T$ is covariant with respect to the phase shifts also, i.e.
\bet
e^{i\theta N}E^T([0,\infty)\times B)e^{-i\theta N} = E^T([0,\infty)\times(B+\theta))
\eeqt
for all $\theta\in [0,2\pi)$ and $B\in\h B([0,2\pi))$, where $\R^2 = [0,\infty)\times [0,2\pi)$ and the sum $B+\theta$ is
understood modulo $2\pi$. (cf. \cite{PellonpääII}).

Since $(q,p)\mapsto W(q,p)$ is a strongly continuous projective representation, and $\R^2$ is connected, Proposition \ref{projections} tells us that
the range of $E^T$ does not contain nontrivial projections. In particular,
the corresponding quantization scheme cannot then produce spectral measures.

\

In this Section, we inspect the possibility of applying the quantization scheme described earlier to the classical position and
momentum variables, using the quantization map $\Gamma^T= L(\cdot,E^T)$ with various generating operators $T$.
For each $k\in\N$, let $x^k$ and $y^k$ denote the functions $(q,p)\mapsto q^k$ and $(q,p)\mapsto p^k$.

The essential question is whether the operator measures $B\mapsto E^T(B\times \R)$ and $B\mapsto E^T(\R\times B)$  are uniquely
determined by their respective moment operator sets $\{\Gamma^T(x^k)\mid k\in \N\}$ and $\{\Gamma^T(y^k)\mid k\in\N\}$.
It is known that this is indeed the case when $T$ is a number state $|n\ket\bra n|$, $n\in\N$ (see \cite{LahtiIII}).
In that case, the operator sets
$\{ \Gamma^T(x^k)\mid k\in\N\}$ and $\{ \Gamma^T(y^k)\mid k\in\N\}$ are eligible to represent the quantizations of $x$ and $y$, respectively.
As is well known, the associated quantum mechanical observables are unsharp position and momentum observables.

Our goal here is to explicitly determine the operators $\Gamma^T(x^k)$ and $\Gamma^T(y^k)$ for certain generating operators $T$.

To begin with, we consider the square integrability domains.
According to Proposition \ref{densityprop} and the associated Remark, these sets are either dense
or trivial. The following two Propositions specify them completely.
 
\begin{proposition}\label{prop5} Let $k\in \N$, let $\eta\in\hil$ be a unit vector, and denote $u=U^{-1}\eta\in L^2(\R)$.
\begin{itemize}
\item[(a)] $\intsqd{x^k}{E^{|\eta\ket\bra\eta|}}\neq \{0\}$ if and only if $\eta\in D(Q^k)$, and in this case,
$\intsqd{x^k}{E^{|\eta\ket\bra\eta|}}=D(Q^k)$.
\item[(b)] The statement of (a) holds true, if ''$x$'' and ''$Q$'' are replaced by ''$y$'' and ''$P$''.
\end{itemize}
\end{proposition}
\begin{proof}
Let $0\neq \vp\in\hil$ and $f=U^{-1}\vp\in L^2(\R)$. We get
\beat
\int_{\R^2} q^{2k} dE^{|\eta\ket\bra\eta|}_{\vp,\vp}(q,p)
&=& \frac {1}{2\pi} \int q^{2k}\left(\int|\bra \vp|W(-q,p)|\eta\ket|^2dp\right)dq\\
&=& \int q^{2k}\left(\int|F(\cc{u}(\cdot-q)f)(p)|^2dp\right)dq\\
&=& \int q^{2k}\left(\int|u(t-q)|^2 |f(t)|^2dt\right)dq\\
&=& \int\left(\int q^{2k}|u(t-q)|^2 |f(t)|^2dq\right)dt\\
&=& \int\int(t-q)^{2k}|u(q)|^2|f(t)|^2 dqdt,
\eeqat
where Lemma 2 of \cite{moments}, the unitarity of the Fourier-Plancherel operator, and Fubini's
theorem have been used. (Since all the functions and measures involved are positive, the calculation is valid regardless
of whether the integrals are finite or not.)

Now the last integral is finite if and only if $\vp$ and $\eta$ are both in $D(Q^k)$. This is seen as follows.

Assume first that the last integral is finite. Then it follows from Fubini's theorem that
$t\mapsto (t-q)^{2k}|f(t)|^2|u(q)|^2$ is integrable for almost all $q$, and $q\mapsto (t-q)^{2k}|u(q)|^2|f(t)|^2$ is integrable
for almost all $t$. Thus $t\mapsto t^{2k}|f(t)|^2$ and $q\mapsto q^{2k}|u(q)|^2$ are integrable.
(The fact that $t\mapsto t^{2k}|f(t)|^2$ is integrable is seen as follows: Take $q\in \R$, such that $|u(q)|^2>0$ and
$t\mapsto (t-q)^{2k}|f(t)|^2|u(q)|^2$ is integrable. This is possible, since $\|\eta\| >0$, which implies that $|u(q)|^2>0$
in some non-null set. Then use the fact that there exist positive constants $A, B, M$, such that
$A t^{2k}\leq (t-q)^{2k}\leq B t^{2k}$ for $|t|\geq M$. The fact that $q\mapsto q^{2k}|u(q)|^2$ is integrable follows
similarly, since we assumed that also $\|\vp\|>0$.) Thus $f$ and $u$ are in the domain of the $k$th power of
the position operator in $L^2(\R)$, so $\vp,\eta\in D(Q^k)$.

Conversely, assume that $\vp,\eta\in D(Q^k)$, so that $t\mapsto t^{2k}|f(t)|^2$ and $q\mapsto q^{2k}|u(q)|^2$ are
integrable. Hence also $t\mapsto |t^{l}||f(t)|^2$ and $q\mapsto |q^l||u(q)|^2$ are integrable for all $l \leq 2k$, implying
that $(t,q)\mapsto (t-q)^{2k}|u(q)|^2|f(t)|^2$ is integrable over $\R^2$. Thus the last integral of the above calculation is
finite.

We conclude that $\intsqd{x^k}{E^{\eta}}\neq \{0\}$ if and only if $\eta\in D(Q^k)$, and in this case,
$\intsqd{x^k}{E^{|\eta\ket\bra\eta|}}=D(Q^k)$.

The result concerning $\intsqd{y^k}{E^{|\eta\ket}}$ is obtained in an analogous manner by using the calculation
\beat
\int_{\R^2} p^{2k} dE^s_{\vp,\vp}(q,p)
&=& \frac {1}{2\pi} \int p^{2k}\left(\int|\bra \vp|W(-q,p)|\eta\ket|^2dq\right)dp\\
&=& \int p^{2k}\left(\int|F^{-1}(\cc{Fu}(\cdot-p)Ff)(q)|^2dq\right)dp\\
&=& \int p^{2k}\left(\int|Fu(t-p)|^2 |Ff(t)|^2dx\right)dp\\
&=& \int\left(\int p^{2k}|Fu(t-p)|^2 |Ff(t)|^2dp\right)dt\\
&=& \int\int(t-p)^{2k}|Fu(p)|^2|Ff(t)|^2 dpdt,
\eeqat
as well as the fact that $P= UF^{-1}U^{-1}QUFU^{-1}$.
\end{proof}

Now we consider the case of an arbitrary positive operator $T$ of trace one.
The following elementary fact is needed. The proof is included for the reader's convenience.

\begin{lemma} Let $T$ be a positive operator of trace one. Let $(\eta_n)$ be an orthonormal sequence
and $(w_n)$ a sequence of nonnegative numbers, such that $T=\sum_n w_n |\eta_n\ket\bra\eta_n|$.
Let $A$ be a closed operator. Then
\bet
\sum_{n=1}^\infty w_n \|A\eta_n\|^2<\infty,
\eeqt
if and only if $A\sqrt{T}$ is a Hilbert-Schmidt operator.
(Here we have denoted $\|A\eta_n\| = \infty$ whenever $\eta_n\notin D(A)$, and used the convention $0\cdot \infty = 0$.)
In particular, the convergence of the series is not dependent on the representation of $T$ in terms of $(\eta_n)$
and $(w_n)$.
\end{lemma}
\begin{proof}
Let $S = \sum_{n=1}^\infty w_n \|A\eta_n\|^2(\leq \infty)$. Assume first that $S<\infty$, so that,
in particular, $\eta_n\in D(A)$ for all those $n\in\N$ for which $w_n>0$. Let $\vp\in\hil$. Since
the series $\sqrt{T} = \sum_n \sqrt{w_n} |\eta_n\ket\bra\eta_n|$ converges in the operator norm,
the vector series $\sum_n \sqrt{w_n} \bra \eta_n |\vp\ket \eta_n$ converges to $\sqrt{T}\vp$ in the norm of
$\hil$. Since $(\eta_n)$ is orthonormal, the Cauchy-Schwartz inequality gives
\bet
\sum_n \sqrt{w_n} |\bra \eta_n |\vp\ket| \|A\eta_n\|\leq \sqrt{S} \|\vp\|<\infty,
\eeqt
so also the series $\sum_n \sqrt{w_n} \bra \eta_n |\vp\ket A\eta_n$ converges in norm. Since $A$ is closed, it follows
that $\sqrt{T}\vp\in D(A)$ and $A\sqrt{T}\vp$ equals the sum of the latter series. In particular, $D(A\sqrt{T}) =\hil$.
Now the previous inequality shows that $\|A\sqrt{T}\vp\|\leq \sqrt{S} \|\vp\|$, so $A\sqrt{T}$ is bounded.
Clearly $\sum_{\xi\in\hK} \|A\sqrt{T}\xi\|^2 = S<\infty$ if $\hK$ is an orthonormal basis of $\hil$ which includes all
the $\eta_n$, so $A\sqrt{T}$ is Hilbert-Schmidt.

Assume then that $A\sqrt{T}$ is a Hilbert-Schmidt operator.
Now $\eta_n=w_n^{-\frac 12}\sqrt{T}\eta_n\in D(A)$ if $w_n>0$, and $S = \sum_{\xi\in\hK} \|A\sqrt{T}\xi\|^2 <\infty$,
where $\hK$ is an orthonormal basis including all the $\eta_n$.
\end{proof}

\begin{proposition}\label{sqdprop}
\begin{itemize}
\item[(a)] Let $k\in \N$. Then $\intsqd{x^k}{E^T} \neq \{0\}$ if and only if $Q^k\sqrt{T}$ is a Hilbert-Schmidt operator, and
in that case, $\intsqd{x^k}{E^T}=D(Q^k)$.
\item[(b)] The statement in (a) holds true, if ''$x$'' and ''$Q$'' are replaced by ''$y$'' and ''$P$''.
\end{itemize}
\end{proposition}
\begin{proof} Write $T$ in the form
$T= \sum_{n=1}^\infty w_n |\eta_n\ket\bra\eta_n|$, where
$\sum_n w_n = 1$, $t_n\geq 0$, and $(\eta_n)$ is an orthonormal sequence in $\hil$. The series converges in the
trace norm, as well as in the operator norm.

For each $\vp\in\hil$, let $A_{\vp,\vp}^{\eta}$ be the density function of the positive measure
$E_{\vp,\vp}^{|\eta\ket\bra\eta|}$. Since the density function of the measure $E^{T}_{\vp,\vp}$ is
$\sum_n w_n A_{\vp,\vp}^{\eta_n}$, we have $\vp\in\intsqd{x^k}{E^T}$ if and only if the function
$x^{2k}\sum_n w_n A_{\vp,\vp}^{\eta_n}$ is integrable over $\R^2$.
In view of the Proposition \ref{prop5}, it is therefore clear that
$\intsqd{x^k}{E^T} \neq \{0\}$ only if $\eta_n\in D(Q^k)$ for all those $n\in \N$ for which $w_n>0$, and that in any case,
$\intsqd{x^k}{E^T}\subset D(Q^k)$.

Assume now that $\eta_n\in D(Q^k)$ for all $n\in\N$ with $w_n>0$, and $0\neq\vp\in D(Q^k)$. Let $u_n = U^{-1}\eta_n$
for each $n$. The monotone convergence theorem and the proof of Proposition \ref{prop5} imply that
\bet
\int x^{2k} dE_{\vp,\vp}^T = \sum_n w_n \int x^{2k} dE_{\vp,\vp}^{\eta_n}=
\int\int\sum_n (t-q)^{2k}w_n|u_n(q)|^2|(U^{-1}\vp)(t)|^2 dtdq
\eeqt
(regardless of whether the series converges or not).
Now if the above integral is finite (i.e. $\vp\in \intsqd{x^k}{E^T}$), then Fubini's theorem gives that
$q\mapsto (t-q)^{2k}|(U^{-1}\vp)(t)|^2\sum_n w_n|u_n(q)|^2$
is integrable for almost all $t$, so by the argument similar to that used in the proof of the preceding Proposition,
it follows that $\sum_n w_n \|Q^k \eta_n\|^2= \int \sum_n w_n q^{2k} |u_n(q)|^2dq<\infty$. On the other hand, if
$\sum_n w_n \|Q^k \eta_n\|^2<\infty$, then each function $q\mapsto \sum_n w_n q^l |u_n(q)|^2$, with $l\leq 2k$ is
integrable, so that $(q,t)\mapsto \sum_n (t-q)^{2k}w_n|u_n(q)|^2|(U^{-1}\vp)(t)|^2$ is integrable over $\R^2$.
(Note that since we assumed that $\vp\in D(Q^k)$, the function $t\mapsto t^l |(U^{-1}\vp)(t)|^2$ is integrable for each
$l\leq 2k$.) Thus $\int x^{2k} dE_{\vp,\vp}^T<\infty$, so $\vp\in \intsqd{x^k}{E^T}$.

We have proved that $\intsqd{x^k}{E^T} \neq \{0\}$ if and only if $\eta_n\in D(Q^k)$ for all $n\in\N$ with $w_n>0$, and
\bet
\sum_{n=1}^\infty w_n \|Q^k\eta_n\|^2<\infty
\eeqt
(where it is understood that $\|Q^k\eta_n\|=\infty$ if $\eta_n\notin D(Q^k)$ and we use the convention $0\cdot\infty =\infty$).
Since $Q^k$ is closed, (a) follows from the preceding Lemma. The statement of (b) is proved similarly, since
\bet
\int x^{2k} dE_{\vp,\vp}^T = \int \int \sum_n w_n (t-p)^{2k}|Fu_n(p)|^2|FU^{-1}\vp(t)|^2 dpdt,
\eeqt
$P = UF^{-1}U^{-1}QUFU^{-1}$, and $P^k$ is also closed. The proof is complete.
\end{proof}

Now we proceed to determine the operators $L(x^k, E^T)$ and $L(y^k,E^T)$ for $T=\sum_n w_n |\eta_n\ket\bra\eta_n|$
satisfying the condition of the preceding Proposition.

\begin{theorem}\label{momentopstheorem}
\begin{itemize}
\item[(a)] Assume that $T$ satisfies the condition of the previous Proposition (a). Then
$L(x^k, E^T) = \sum_{l=0}^k s^Q_{kl} Q^l$, where $s^Q_{kl} = \binom {k}{l} (-1)^{k-l} \tr[Q^{k-l}T]$, with
each $Q^{k-l}T$ a trace class operator.
\item[(b)] The statement in (a) holds true, if ''(a)'', ''$x$'' and ''$Q$'' are replaced by ''(b)'', ''$y$'' and ''$P$''.
\end{itemize}
\end{theorem}
\begin{proof} Assume first that $T=|\eta\ket\bra\eta|$ for $\eta\in D(Q^k)$. Denote $u=U^{-1}\eta\in L^2(\R)$.
Define a polynomial $p^\eta:\R\to \R$ by
\bet
p^\eta(t) = \bra \eta|(t-Q)^k\eta\ket=\sum_{l=0}^k\binom kl (-1)^{k-l}\bra \eta|Q^{k-l}\eta\ket t^l.
\eeqt
Since $p^\eta$ is a polynomial of order $k$, the operator $p^\eta(Q)$ is selfadjoint, and has the
domain $D(Q^k)$. Thus by Proposition \ref{prop5}, we have $D(p^\eta(Q))=D(Q^k) = \intsqd{x^k}{E^{|\eta\ket\bra\eta|}}$.

Let $\vp\in \intsqd{x^k}{E^{|\eta\ket\bra\eta|}}\subset \intd{x^k}{E^{|\eta\ket\bra\eta|}}$, and $\psi\in\hil$.
Let $f=U^{-1}\vp$, $g=U^{-1}\psi$. Since the function
\bet
(q,p)\mapsto q^k \bra \psi|W(-q,p)|\eta\ket\cc{\bra \vp|W(-q,p)|\eta\ket}
\eeqt
is integrable over $\R^2$ (by the definition of $\intd{x^k}{E^{|\eta\ket\bra\eta|}}$), we get
\beat
\bra \psi|L(x^k, E^{|\eta\ket\bra\eta|})\vp\ket &=& \int_{\R^2} q^{k} dE^{|\eta\ket\bra\eta|}_{\psi,\vp}(q,p)\\
&=& \frac {1}{2\pi} \int q^{k}\left(\int\bra \psi|W(-q,p)|\eta\ket\cc{\bra \vp|W(-q,p)|\eta\ket} dp\right)dq\\
&=& \int q^{k}\left(\int \cc{F(\cc{u}(\cdot-q)g)(p)}F(\cc{u}(\cdot-q)f)(p) dp\right)dq\\
&=& \int q^{k}\left(\int \cc{u(t-q)}\cc{g(t)} u(t-q)f(t)dt\right)dq\\
&=& \int\left(\int q^{k}|u(t-q)|^2 dq\right)\cc{g(t)}f(t)dt\\
&=& \int\left(\int(t-q)^{k}|u(q)|^2 dq\right) \cc{g(t)}f(t)dt,\\
&=& \int\ \bra \eta| (t-Q)^k |\eta\ket\cc{g(t)}f(t)dt\\
&=& \bra \psi|p^\eta(Q)\vp\ket.
\eeqat
The fifth equality follows from Fubini's theorem, since $(q,t)\mapsto q^{k}|u(t-q)|^2\cc{g(t)}f(t)$ is integrable
(because of the Cauchy-Schwarz inequality and the square integrability of the maps
$(q,t)\mapsto |u(t-q)g(t)|$ and $(q,t)\mapsto |q^k u(t-q) f(t)|$, the latter being a consequence of the
proof of Proposition \ref{prop5}.)
It follows that $p^\eta(Q)\subset L(x^k, E^{|\eta\ket\bra\eta|})$.

The equality $p^{\eta}(Q) = L(x^k, E^{|\eta\ket\bra\eta|})$ follows from the fact that being selfadjoint, the operator
$p^\eta(Q)$ cannot have a proper symmetric extension.

Now we take $T=\sum_n w_n |\eta_n\ket\bra\eta_n|$ under the condition of the preceding Proposition, so that
$\intsqd{x^k}{E^T} = D(Q^k)=\intsqd{x^k}{E^{|\eta_n\ket\bra\eta_n|}}$ for each $n$. Let $\vp\in D(Q^k)$ and
$\psi\in\hil$, and $f, g$ be as before.

According to Proposition 1 of \cite{moments}, we have
\beat
\bra \psi|L(x^k, E^T)\vp\ket &=& \sum_{n=1}^\infty w_n\bra\psi|L(x^k, E^{|\eta_n\ket\bra\eta_n|})\vp\ket
= \sum_{n=1}^\infty w_n \bra \psi |p^{\eta_n}(Q)\vp\ket\\
&=& \sum_{n=1}^\infty w_n \sum_{l=0}^k\binom kl (-1)^{k-l}\bra \eta_n|Q^{k-l}\eta_n\ket \bra \psi|Q^l\vp\ket.
\eeqat
Since
\bet
\|Q^{k-l}\eta_n\|^2 = \int q^{2(k-l)}|(U^{-1}\eta_n)(q)|^2 dq\leq 1 + \int q^{2k}|(U^{-1}\eta_n)(q)|^2 dq = 1+\|Q^k\eta_n\|,
\eeqt
it follows from the preceding Lemma that also each $Q^{k-l}\sqrt{T}$ is defined in all of $\hil$ and is a Hilbert-Schmidt
operator. Thus each $Q^{k-l}T = Q^{k-l}\sqrt{T}\sqrt{T}$ is defined in all of $\hil$ and is a bounded operator of trace class.
Thus the series $\sum_n w_n \bra \eta_n|Q^{k-l}\eta_n\ket$ converges
(clearly to $\tr[Q^{k-l}T]$) for each $l$, so
\bet
\bra \psi|L(x^k, E^T)\vp\ket = \sum_{l=0}^k s_{kl}^Q \bra \psi |Q^l\vp\ket.
\eeqt
Since $\vp$ was arbitrarily chosen from the set $\intsqd{x^k}{E^T} = D(Q^k)$ and the operator
$\sum_{l=0}^k s_{kl}^QQ^l$ is selfadjoint, we conclude that $L(x^k, E^T)=\sum_{l=0}^k s_{kl}^QQ^l$, and
the proof of (a) is complete.

The statement (b) is proved similarly by using the unitary equivalence of $Q$ and $P$.
\end{proof}
\noindent {\bf Remark.} As mentioned at the beginning of the Section, the uniqueness of the operator measure
which gives the moment operators of Theorem \ref{momentopstheorem}, is verified only in the case where $T=|n\ket\bra n|$
for some $n$. The uniqueness question in the general case remains open.

\ 

We close this Section with a remark on an another application of our quantization scheme. Consider the function
$h(q,p)=\frac 12(q^2+p^2)$, i.e. the classical oscillator energy variable. It is known that for each $n\in\N$, the
operators $\Gamma^{|n\ket\bra n|}(h^k)$, $k\in\N$, are
the moment operators of the polar margin of the phase space observable $E^{|n\ket\bra n|}$, and that the marginal observable
is uniquely determined by its moments \cite{LahtiIV}. Thus a quantization of $h$ is given by the set
$\{ \Gamma^{|n\ket\bra n|}(h^k) \mid k\in \N\}$ of operators. These operators were determined explicitly in \cite{Lahti}; they are
certain polynomials of the usual oscillator Hamiltonian $\frac 12 (Q^2+P^2)$. The quantized oscillator energy observable
is the unsharp number observable (see \cite[p. 90]{OQP}).

\section{Optimal phase space quantization in $\R^2$}

Consider the situation of the previous Section. At least in the case where $T=|n\ket\bra n|$, the quantizations of the position and
momentum variables $x$ and $y$ corresponding to the covariant quantization map $\Gamma^T= L(\cdot, E^T)$ are the Cartesian
margins of $E^T$, or, equivalently, the sets of the operators $\{\Gamma^T(x^k)\mid k\in\N\}$ and $\{\Gamma^T(y^k)\mid k\in \N\}$.
If the margins were projection valued, the quantization of e.g. $x$ would just be the spectral measure of $L(x,E^T)$,
with each operator $L(x^k, E^T)$ equal to the corresponding power of $L(x,E^T)$. Although this is not the case,
we can still try to find those generating operators $T$ for which the situation would be in some sense close to this ideal situation,
where only the first power of $x$ is needed to determine its quantization.

First, we can find the generating operators $T$ which give $L(x,E^T) = Q$ and $L(y,E^T) = P$, so as to make the operators $\Gamma^T(x)$
and $\Gamma^T(y)$ equal to the actual position and momentum operators. In view of Theorem \ref{momentopstheorem}, we know that
the square integrability domains of $L(x, E^T)$ and $L(y, E^T)$ are nontrivial if and only if
$Q\sqrt{T}$ and $P\sqrt{T}$ are Hilbert-Schmidt operators.
In that case we have,
\beat
L(x,E^T)&=&Q-\tr[QT] I;\\
L(y, E^T) &=& P-\tr[PT] I.
\eeqat
So if we assume that the square integrability domains of $L(x,E^T)$ and $L(y, E^T)$ are nontrivial,
we have $L(x, E^T) = Q$ and $L(y, E^T)=P$ exactly when $T$ is such that $\tr[QT]=\tr[PT]=0$.
This occurs, for example, if we choose $T$ to be a mixture of number states, i.e.
$T = \sum_n w_n |n\ket\bra n|$. Then the above Hilbert-Schmidt conditions take the form $\sum_n w_n n<\infty$
\cite{moments}.

Consider next the operators $\Gamma^T(x^2)$ and $\Gamma^T(y^2)$. According to Theorem \ref{momentopstheorem}, they are given by
\beat
L(x^2, E^T) = Q^2-2\tr[QT] Q+\tr[Q^2T]I;\\
L(y^2, E^T) = P^2-2\tr[PT] P+\tr[P^2T]I,
\eeqat
provided that $Q^2\sqrt{T}$ and $P^2\sqrt{T}$ are Hilbert-Schmidt operators (or, equivalently, that
$\intsqd{x^2}{E^T}$ and $\intsqd{y^2}{E^T}$ are nontrivial).
In order to make the situation close to the spectral measure case, we would like to minimize the ''noise'' operators
$R^T(x) = L(x^2, E^T)^2-L(x,E^T)^2$ and $R^T(y) = L(y^2, E^T)^2-L(y,E^T)^2$. Now
\beat
R^T(x) = (\tr[Q^2T]-\tr[QT]^2)I = \text{Var}(Q, T)I;\\
R^T(y) = (\tr[P^2T]-\tr[PT]^2)I = \text{Var}(P, T)I,
\eeqat
on the domains $D(Q^2)$ and $D(P^2)$, respectively, where e.g. $\text{Var} (Q, T)$ denotes the variance of the probability
measure $p^Q_T: = \tr[TE^Q(\cdot)]$, with $E^Q$ the spectral measure of $Q$. The last equalities are obtained as follows:
Let $T=\sum_n w_n |\eta_n\ket\bra\eta_n|$, $p^Q_T= \tr[TE^Q(\cdot)]$, and $p_n^Q=\bra\eta_n|E^Q(\cdot)\eta_n\ket$,
where $E^Q$ is the spectral measure of $Q$. Now $p^Q_T = \sum_n w_n p_n^Q$, with the
series converging absolutely in the total variation norm, so we have (by e.g. Lemma 1 of \cite{moments}) that
\beat
\tr[Q^2T] -\tr[QT]^2 &=& \sum_n w_n \bra \eta_n |Q^2\eta_n\ket - (\sum_n w_n \bra \eta_n |Q\eta_n\ket)^2 \\
&=& \sum_n w_n \int x^2 dp^Q_n - (\sum_n w_n \int x dp^Q_n)^2 = \int x^2 dp^Q_T -(\int xdp^Q_T)^2.
\eeqat
Therefore, $R^T(x) = \text{Var} (Q, T)$. The result $R^T(y) = \text{Var} (P, T)I$ follows similarly.
Since $\text{Var} (Q, T)$ and $\text{Var} (P, T)$ are always positive, we see explicitly that $R^T(x)$ and $R^T(y)$
are never zero.

As is well known, the generating operator $T$ can be chosen so that e.g. $R^T(x)=\text{Var} (Q, T)I$ is arbitrarily small
(in the sense that $\|R^T(x)\| = \text{Var} (Q, T)$ is such), but then
$R^T(y)$ becomes large, because of the inequality $\text{Var} (Q, T)\text{Var} (P, T)\geq \frac 14$.
The product $R^T(x)R^T(y)$ can reach its lower bound $\frac 14$ only in the case where $T$ is a vector state
of minimal uncertainty. If we assume that $\tr[QT]=0=\tr[PT]$ as discussed before, the
operators $T$ that give $R^T(x)R^T(y) = \frac 14$ are of the form $T=|\eta\ket\bra\eta|$, with
\bet
(U^{-1}\eta)(t) = (\sqrt{\pi}\Delta q)^{-\frac 12}e^{-\frac {t^2}{4(\Delta q)^2}},
\eeqt
where $\Delta q>0$ (in fact, $(\Delta q)^2 = \text{Var} (Q, |\eta\ket\bra\eta|)$
\cite[p. 92]{Thirring}. Moreover, we could require that $R^T(x)=R^T(y)$, so as to make the situation
symmetric between $x$ and $y$. This leaves us with only one generating operator, namely $T=|0\ket\bra 0|$.
Note that this choice indeed gives a quantization of position and momentum, for the associated operator measure is uniquely
determined by its moment operators (see Remark of Theorem \ref{momentopstheorem}).

\section{When is a positive operator measure projection valued?}

Let $E:\hB(\R)\to L(\hil)$ be a positive operator measure.
If $E$ is a spectral measure, the first moment $L(x,E)$ is always selfadjoint on the domain $\intsqd{x}{E}$, and
$\int x^2 dE_{\vp,\vp}= \|L(x, E)\vp\|^2$ for all $\vp\in \intsqd{x}{E}$. In the case of a general positive operator measure, this
need not be true, as the above case of the Cartesian margins of the phase space observable $E^T$
demonstrates. It turns out that this condition is sufficient for a positive operator $E$ to be a spectral measure. The proof
of this fact can be found in \cite[p. 130]{Akhiezer}. Since that proof does not contain certain
details which are perhaps not obvious, we give a (slightly different) proof here as part (b) of the following Proposition.

An adaptation of the steps leading to the result in \cite[p. 466]{Riesz}
gives part (a) of the following Proposition. For each $k\in\N$, we let $\tilde{L}(x^k,E)$ denote the restriction of $L(x^k,E)$ to $\intsqd{x^k}{E}$.
\begin{proposition}
Let $E:\hB(\R)\to L(\hil)$ be a positive operator measure, such that
\bet
\int x^2 dE_{\vp,\vp}= \|L(x, E)\vp\|^2
\eeqt
for all $\vp\in \intsqd{x}{E}$.
\begin{itemize}
\item[(a)] $\tilde{L}(x^n,E) = \tilde{L}(x,E)^n$ for all $n\in \N$.
\item[(b)] If $\tilde{L}(x,E)$ is assumed to be selfadjoint, then $E$ is projection valued.
\end{itemize}
\end{proposition}
\begin{proof}
Let $P:\hB(\R)\to \kil$ be a Naimark dilation of $E$ into a spectral measure acting on a Hilbert space $\kil$.
Let $V:\hil\to\kil$ be the associated isometric map, so that $E(B) = V^*P(B)V$ for all $B\in\hB(\R)$. Denote by
$P_{\hil}$ the projection $VV^*$, acting on $\kil$ with $V\hil$ as its range. (Note that $V^*V$ is the identity operator of $\hil$.)
Now $\tilde{L}(x^k,E)=V^*L(x^k,P)V$
for each $k\in\N$ (see \cite{LahtiII}). Since $P$ is a spectral measure, we thus have
\be\label{dilation}
\tilde{L}(x^k,E)=V^*A^kV
\eeq
for all $k\in\N$, where $A=L(x,P)$. Denote $E_1 = \tilde{L}(x,E)$. We prove by induction that for each $n\in\N$,
\be\label{AVLind}
\intsqd{x^n}{E} = D(E_1^n), \text{ and } A^nV\vp = VE_1^n\vp \text{ for all } \vp\in \intsqd{x^n}{E}.
\eeq
Take first $n=1$, and let $\vp\in \intsqd{x}{E}=D(E_1)=D(AV)$. Since the
measures $E_{\vp,\vp}$ and $P_{V\vp,V\vp}$ are the same, and $P$ is a spectral measure, the assumption implies that
\bet
\|AV\vp\|^2 = \int x^2 dP_{V\vp,V\vp} = \int x^2 dE_{\vp,\vp} = \|E_1\vp\|^2.
\eeqt
Using (\ref{dilation}) and the fact that $V$ is isometric, we thus get
\bet
\|AV\vp\|^2 = \|E_1\vp\|^2= \|V^*AV\vp\|^2 = \|P_{\hil}AV\vp\|^2.
\eeqt
Since $P_{\hil}$ is a projection, this means that
\be\label{AVL}
AV\vp = P_{\hil}AV\vp= VE_1\vp \text{ for all } \vp\in \intsqd{x}{E}=D(E_1),
\eeq
i.e. (\ref{AVLind}) holds for $n=1$.
Now let $k\in \N$, $k>1$, and assume that (\ref{AVLind}) holds for $n=k-1$. Let $\vp\in \intsqd{x^k}{E}$. By (\ref{dilation}),
this implies that $V\vp\in D(A^k)$, so that $AV\vp\in D(A^{k-1})$. Since $\vp\in \intsqd{x^k}{E}\subset \intsqd{x}{E}$,
it thus follows from (\ref{AVL}) that $V(E_1\vp)=AV\vp\in D(A^{k-1})$, so (\ref{dilation}) and the induction assumption give
$E_1\vp\in \intsqd{x^{k-1}}{E}=D(E_1^{k-1})$. Hence, $\vp\in D(E_1^k)$. Conversely, if $\vp\in D(E_1^k)$, then
$\vp\in \intsqd{x}{E}=D(AV)$ and $E_1\vp\in D(E_1^{k-1})=\intsqd{x^{k-1}}{E}$, so $AV\vp=V(E_1\vp)\in D(A^{k-1})$ by (\ref{AVL}) and
(\ref{dilation}), implying that $V\vp\in D(A^k)$, i.e. $\vp\in \intsqd{x^k}{E}$. Thus, $\intsqd{x^k}{E}=D(E_1^k)$. Let $\vp$ be in this set.
Since now $E_1\vp\in D(E_1^{k-1})$, the induction assumption (along with the fact that $AV\vp=V(E_1\vp)$) gives
\bet
A^kV\vp = A^{k-1}(AV\vp) = A^{k-1}V(E_1\vp) = VE_1^{k-1}(E_1\vp) = VE_1^k\vp,
\eeqt
completing the induction proof of (\ref{AVLind}).

Let $n\in \N$. Now (\ref{dilation}) and (\ref{AVLind}) give $\tilde{L}(x^n,E)\vp=V^*A^nV\vp = V^*VE_1^n\vp = E_1^n\vp$
for all $\vp\in \intsqd{x^n}{E}=D(E_1^n)$, so $\tilde{L}(x^n,E)=\tilde{L}(x,E)^n$. This proves (a).

If we assume that $\tilde{L}(x,E)$ is selfadjoint, it follows from \eqref{AVL} that $P_{\hil}D(A)\subset D(A)$. This fact is proved in \cite{Szaf},
but we include the proof here for the reader's convenience. To that end, let $\psi\in D(A)$, and let $\vp\in D(E_1)$ be arbitrary.
Using \eqref{AVL}, we get
\bet
\bra E_1\vp|V^*\psi\ket = \bra VE_1\vp |\psi\ket = \bra AV\vp|\psi\ket  = \bra \vp |V^*A\psi\ket,
\eeqt
which implies that $V^*\psi\in D(E_1^*)$. Since $E_1$ is selfadjoint, $V^*\psi\in D(E_1)$, so $P_{\hil}\psi = V(V^*\psi)\in VD(E_1)$.
But $VD(E_1)$ is contained in $D(A)$, because $D(E_1) = D(AV)$. Thus $P_{\hil}\psi\in D(A)$, proving the fact $P_{\hil}D(A)\subset D(A)$.
In addition, the above calculation shows that $V^*D(A)\subset D(E_1)$, and $E_1V^*\psi = E_1^*(V^*\psi) = V^*A\psi$ for all $\psi\in D(A)$.
Combining this with \eqref{AVL}, we get
\bet
P_{\hil}A\psi= VV^*A\psi = V(E_1V^*\psi) = VE_1(V^*\psi) = AV(V^*\psi) = AP_{\hil}\psi
\eeqt
for all $\psi\in D(A)$. Consequently, $P_{\hil}A\subset AP_{\hil}$. Since $A$ is selfadjoint, this implies that $P_{\hil}$ commutes
with all the spectral projections $P(B)$ \cite[pp. 320, 301]{Riesz}. It follows that each $E(B)$ is a projection \cite[Corollary 2.2.2.]{LahtiV},
so the proof is complete.
\end{proof}

\noindent{\bf Remark. } As mentioned before, the result appearing in part (b) of the above Proposition can be found in the classic book of Akhiezer and Glazman
\cite{Akhiezer}. The result seems to be somewhat well known (see e.g. \cite{McKelvey, Wan}, both of which refer to the works of Akhiezer and Glazman).
However, the fact is given in a much later work \cite{Kruszynski}
without reference to \cite{Akhiezer} (though we have not been able to convince ourselves of their argumentation), and R. Werner \cite[p. 796]{Werner} only
mentions that it holds for normalized
compactly supported operator measures. Moreover, Ingarden \cite[p. 87]{Ingarden} says that all the semispectral measures with the same selfadjoint first moment $A$,
have variances greater than or equal to that of the spectral measure of $A$. Part (b) of the above Proposition gives more - it asserts that the minimum variance
occurs \emph{only} in the case of the spectral measure of $A$. We note also that the proof given in \cite[p. 87]{Ingarden} considers only compactly supported
semispectral measures, and contains no reference to \cite{Akhiezer}.

\

Now we get the following characterization for projection valued measures.

\begin{theorem} Let $E:\hB(\R)\to L(\hil)$ be a positive normalized operator measure, such that $\tilde{L}(x,E)$ is selfadjoint. Then
the following conditions are equivalent.
\begin{itemize}
\item[(i)] E is a spectral measure;
\item[(ii)] $L(x^2,E)=L(x,E)^2$;
\item[(iii)] $\int x^2 dE_{\vp,\vp}= \|L(x, E)\vp\|^2$ for all $\vp\in \intsqd{x}{E}$.
\end{itemize}
\end{theorem}
\begin{proof} Since $\tilde{L}(x,E)$ is selfadjoint, it coincides with its symmetric extension $L(x,E)$.

Assume that (i) holds. Then $\tilde{L}(x^2,E)=\tilde{L}(x,E)^2$ by a standard result of spectral theory. Since
$E$ is projection valued, we have also $\tilde{L}(x^2,E)= L(x^2,E)$, so (ii) holds.

Assume (ii). Then we have
\be\label{equality}
\int x^2 dE_{\vp,\vp} = \|L(x,E)\vp\|^2
\eeq
for all $\vp\in D(x^2,E)$. In order to get (iii), we have to establish this identity for vectors in the larger set $D(x,E)=\intsqd{x}{E}$.
Let $\vp\in \intsqd{x}{E}= D(x,E)$. Since $L(x,E) = \tilde{L}(x,E)$
is selfadjoint, the closure of the restriction of $L(x,E)$ to the domain
of $L(x,E)^2$ is $L(x,E)$ itself \cite[p. 1245]{DunfordII}. Therefore, by (ii), we can pick a sequence $(\vp_n)$ of vectors in $D(x^2,E)$, converging
to $\vp$, such that $(L(x,E)\vp_n)$ converges to $L(x,E)\vp$. Since $\vp_n\in D(x^2,E)$ for each $n$, \eqref{equality} gives that
\bet
\lim_n \int x^2 dE_{\vp_n,\vp_n} = \lim_n \|L(x,E)\vp_n\|^2 = \|L(x,E)\vp\|^2.
\eeqt
Since $|E_{\vp_n,\vp_n}(B)-E_{\vp,\vp}(B)|\leq (\|\vp_n\|+\|\vp\|)\|E(\R)\|\|\vp_n-\vp\|$ for all $n\in \N$ and $B\in \hB(\R)$, the sequence
$(E_{\vp,\vp}(B))$ converges to $E_{\vp,\vp}(B)$ uniformly for $B\in \hB(\R)$, so the sequence 
$(E_{\vp,\vp})$ of positive measures converges to $E_{\vp,\vp}$ in the total variation norm \cite[p. 97]{Dunford}. It follows by
\cite[Lemma A.5]{Lahti} that
\be\label{eqII}
\int x^2 dE_{\vp,\vp} \leq \|L(x,E)\vp\|^2, \ \ \ \vp\in D(x,E).
\eeq

It follows e.g. from the proof of Lemma A.2 of \cite{Lahti} (see \cite[p. 65]{Beukema}) that $\|L(x,E)\vp\|^2\leq \int x^2 dE_{\vp,\vp}$
for all $\vp\in D(x,E)$.
Combining this with \eqref{eqII}, we get (iii).

Because (iii) implies (i) by the preceding Proposition, the proof is complete.
\end{proof}

\end{document}